# Secure multiplex coding to attain the channel capacity in wiretap channels

Daisuke Kobayashi

NTT Data Corporation

3-2-18 Shiba, Minato-ku, Tokyo, 105-0014 Japan

Email: kobayashidib@nttdata.co.jp

Hirosuke Yamamoto

Graduate School of Frontier Sciences, University of Tokyo

5-1-5 Kashiwanoha, Kashiwa-shi, Chiba, 277–8561 Japan

Email: Hirosuke@ieee.org

Tomohiro Ogawa

Graduate School of Information Science and Technology, University of Tokyo

7-3-1 Hongo, Bunkyo-ku, Tokyo, 113-8656 Japan

Email: ogawa@mist.i.u-tokyo.ac.jp

## Abstract

It is known that a message can be transmitted safely against any wiretapper via a noisy channel without a secret key if the coding rate is less than the so-called secrecy capacity $C_S$, which is usually smaller than the channel capacity $C$. In order to remove the loss $C - C_S$, we propose a multiplex coding scheme with plural independent messages. In this paper, it is shown that the proposed multiplex coding scheme can attain the channel capacity as the total rate of the plural messages and the perfect secrecy for each message. The coding theorem is proved by extending Hayashi's proof, in which the coding of the channel resolvability is applied to wiretap channels.

## I. INTRODUCTION

When Alice sends a message to Bob via a public noisy channel, Eve may wiretap the message. But, since the main channel from Alice to Bob has usually a different characteristic from the wiretap channel from Alice to Eve, we can devise a code such that the perfect secrecy against Eve can be attained without any secret key. The maximum attainable coding rate of such codes is called the *secrecy capacity* $C_S$, which is generally smaller than the channel capacity $C$ of the main channel.

The coding problem for the wiretap channels was first studied by Wyner [2]. Although the main and wiretap channels can be considered as a kind of broadcast channel [1], Wyner proved the coding theorem for the case of the so-called degraded broadcast channel. For more general broadcast channels, Csiszár and Körner [3] proved that the secrecy capacity $C_S$ is given by $C_S = \max_{\widetilde{X} \to X \to (Y,Z)} [I(\widetilde{X};Y) - I(\widetilde{X};Z)]$, where $X$ is the input, and $Y$ and $Z$ are the output of the main and wiretap channels, respectively. $\widetilde{X}$ is an auxiliary random variable that makes a Markov chain $\widetilde{X} \to X \to (Y,Z)$. In order to achieve the perfect secrecy with a positive coding rate, the channels must satisfy that $I(\widetilde{X};Y) > I(\widetilde{X};Z)$ for some $\widetilde{X}$, i.e, the main channel must be less noisy than the wiretap channel in this sense. But, even in the case that the main channel is more noisy than the wiretap channel, Maurer [4] devised a protocol which can attain the perfect secrecy if a public noiseless channel can be used.

In the above studies, channels are assumed to be stationary. On the contrary, the information-spectrum methods [5] have been developed recently, and many kinds of coding problems have been proved for the so-called general sources and general channels, which might be neither Ergodic nor stationary. As one of them, Han and Verdú [6] studied the so-called channel resolvability problem, in which we want to approximate the output probability distribution of a noisy channel for a given input probability distribution by encoding a random number. The minimum rate of the random number necessary to attain the approximation is called the channel resolvability, and they developed the theory of the channel resolvability for general channels. On the other hand, for quantum channels, Devetak [8] introduced a stochastic encoder to realize a non-distinguishable probability distribution for any message at the output of the wiretap channel. Based on these background, Hayashi [9] considered the coding problem of general wiretap channels in the framework of the stochastic encoders and the channel resolvability, and established the proving method for the coding theorem of general wiretap channels.

It is well known from the coding theorems proved in the previous studies that messages cannot be transmitted at any rate larger than the secrecy capacity $C_S$ if we want to attain the perfect secrecy. Since $C_S$ is generally less than the channel capacity $C$, we must loss $C - C_S$ in exchange for the secrecy. But, in this paper, we will devise a multiplex coding of plural independent messages to remove the loss, and we will show, according to Hayashi's proving method, that the channel capacity $C$ can be attained as the total rate of the plural messages and each message can be protected with the perfect secrecy.

In section 2, we define several technical terms, which are used in the information-spectrum method.

The multiplex coding scheme of plural messages is proposed in section 3. The main theorem is also shown in section 3 although it is proved in section 4. Finally, the case of stationary memoryless wiretap channels is treated in section 5. In this paper, both input and output alphabets of channels are assumed to be discrete.

## II. PRELIMINARIES

According to the information-spectrum method [5], we represent a general random process, which might be neither Ergodic nor stationary, as

$$\mathbf{X} = \{X^n\}_{n=1}^{\infty}, \tag{1}$$

where $X^n = (X_1, X_2, \cdots, X_n)$ and each $X_i$ take values in discrete alphabet $\mathcal{X}^n$ and $\mathcal{X}$, respectively, and $\mathcal{X}^n$ is the Cartesian product of $\mathcal{X}$.

For two general random process $\mathbf{X}$ and $\mathbf{Y}$, the spectral sup-mutual information rate and the spectral inf-mutual information rate are defined as follows.

**Definition 1**

Spectral sup-mutual information rate :

$$\overline{I}(P_\mathbf{X}, P_{\mathbf{Y}|\mathbf{X}}) \equiv \inf \left\{ \alpha \mid \lim_{n \to \infty} \Pr \left\{ \frac{1}{n} \log \frac{P_{Y^n|X^n}(Y^n|X^n)}{P_{Y^n}(Y^n)} > \alpha \right\} = 0 \right\} \tag{2}$$

Spectral inf-mutual information rate :

$$\underline{I}(P_\mathbf{X}, P_{\mathbf{Y}|\mathbf{X}}) \equiv \sup \left\{ \beta \mid \lim_{n \to \infty} \Pr \left\{ \frac{1}{n} \log \frac{P_{Y^n|X^n}(Y^n|X^n)}{P_{Y^n}(Y^n)} < \beta \right\} = 0 \right\}$$

In the case that $\mathbf{X}$ and $\mathbf{Y}$ are i.i.d. processes with probability distribution $P_{X,Y}$, both of the spectral sup- and inf-mutual information rates coincide with the ordinary mutual information $I(X;Y)$.

A general channel $\mathbf{W}$ with an input alphabet $\mathcal{X}$ and an output alphabet $\mathcal{Y}$ is defined as $\mathbf{W} = \{W^n\}_{n=1}^{\infty}$, where $W^n = W^n(\cdot|\cdot)$ is an arbitrary conditional probability distribution that satisfies

$$\sum_{y^n \in \mathcal{Y}^n} W^n(y^n|x^n) = 1 \tag{3}$$

for each $x^n \in \mathcal{X}^n$ and each $n = 1, 2, \cdots$. For the input process $\mathbf{X} = \{X^n\}_{n=1}^{\infty}$ and the output process $\mathbf{Y} = \{Y^n\}_{n=1}^{\infty}$ of the general channel $\mathbf{W}$, $W^n$ satisfies for any $n > 0$ that

$$P_{X^n,Y^n}(x^n, y^n) = P_{X^n}(x^n) W^n(y^n|x^n), \tag{4}$$

$$P_{Y^n|X^n}(y^n|x^n) = W^n(y^n|x^n), \tag{5}$$

$$P_{Y^n}(y^n) = \sum_{x^n} P_{X^n}(x^n) W^n(y^n|x^n)$$

$$\equiv P_{X^n} W^n(y^n). \tag{6}$$

Note that $P_{Y^n}$ is also represented as $P_{X^n} W^n$ because $P_{Y^n}$ is determined by $P_{X^n}$ and $W^n$.

For simplicity, we represent the alphabets of a general channel as $\mathcal{X} \to \mathcal{Y}$ when the input and output alphabets are $\mathcal{X}$ and $\mathcal{Y}$, respectively. Let $\mathbf{U}$ and $\mathbf{W}$ be general channels with alphabets $\widetilde{\mathcal{X}} \to \mathcal{X}$ and $\mathcal{X} \to \mathcal{Y}$, respectively. Then, the cascade channel $\mathbf{UW}$ with alphabets $\widetilde{\mathcal{X}} \to \mathcal{Y}$ is defined by $\mathbf{UW} = \{(UW)^n\}_{n=1}^\infty$, where

$$(UW)^n(y^n|\widetilde{x}^n) \equiv \sum_{x^n} U(x^n|\widetilde{x}^n)W(y^n|x^n). \tag{7}$$

Now we consider the channel resolvability problem. Let $\mathbf{V} = \{V^n\}_{n=1}^\infty$ be a given general channel with alphabets $\mathcal{X} \to \mathcal{Z}$, an input $\mathbf{X} = \{X^n\}_{n=1}^\infty$, and the corresponding output $\mathbf{Z} = \{Z^n\}_{n=1}^\infty$. Then, we want to approximate the output $\mathbf{Z}$ by inputting $\widehat{\mathbf{X}} = \{\widehat{X}^n\}_{n=1}^\infty$ into the channel, where $\widehat{X}^n$ is generated by encoding a uniform random number $K$ over an alphabet $\mathcal{K} \equiv \{1, 2, \cdots, M_n\}$. For the output $\widehat{\mathbf{Z}} = \{\widehat{Z}^n\}_{n=1}^\infty$ of the input $\widehat{\mathbf{X}}$, we evaluate the performance of the approximation between $\mathbf{Z}$ and $\widehat{\mathbf{Z}}$ by the variational distance $d(Z^n, \widehat{Z}^n) = ||P_{Z^n} - P_{\widehat{Z}^n}||_1 = \sum_{z^n} |P_{Z^n}(z^n) - P_{\widehat{Z}^n}(z^n)|$.

**Definition 2**

For a given channel $\mathbf{V} = \{V^n\}_{n=1}^\infty$ with alphabets $\mathcal{X} \to \mathcal{Y}$, a rate $R$ is called achievable for an input $\mathbf{X}$ if there exists a sequence of codes $\varphi_n : \widehat{X}^n = \varphi_n(K)$ that satisfies

$$\lim_{n \to \infty} d(Z^n, \widehat{Z}^n) = 0, \tag{8}$$

$$\limsup_{n \to \infty} \frac{1}{n} \log M_n \leq R, \tag{9}$$

where $K$ is the uniform random number over $\mathcal{K} = \{1, 2, \cdots, M_n\}$, and $P_{\widehat{Z}^n}(z^n) = P_{\widehat{X}^n} V^n(z^n)$. Furthermore, the channel resolvability for the input $\mathbf{X}$, denoted by $S_\mathbf{X}(\mathbf{V})$, is defined as follows.

$$S_\mathbf{X}(\mathbf{V}) = \inf\{R \,|\, R \text{ is achievable for the input } \mathbf{X} \text{ of the channel } \mathbf{V}\} \tag{10}$$

Then, the next theorem is proved by Han-Verdú [6].

**Theorem 1** For any general channel $\mathbf{V}$ and any input $P_\mathbf{X}$, it holds that

$$S_\mathbf{X}(\mathbf{V}) \leq \overline{I}(P_\mathbf{X}, \mathbf{V}). \tag{11}$$

Furthermore, it is shown in [7] that if $\mathbf{V}$ is a full-rank channel, i.e. $\mathbf{V}$ is a stationary memoryless channel such that $\{V(\cdot|x)\}$, $x \in \mathcal{X}$, are independent as a set of vector, then the channel resolvability $S_\mathbf{X}(\mathbf{V})$ satisfies that for any input $\mathbf{X}$,

$$S_\mathbf{X}(\mathbf{V}) = \overline{I}(P_\mathbf{X}, \mathbf{V}). \tag{12}$$

## III. MULTIPLEX CODING

In this section, we consider a multiplex coding for wiretap channels. Assume that Alice sends messages to Bob via a main channel $\mathbf{W}$ with alphabets $\mathcal{X} \to \mathcal{Y}$ and Eve eavesdrops the messages via a wiretap channel $\mathbf{V}$ with alphabets $\mathcal{X} \to \mathcal{Z}$. $\mathbf{W}$ and $\mathbf{V}$ are general channels in the sense of the information-spectral method. The input of both channels is $P_{\mathbf{X}}$, and the outputs of $\mathbf{W}$ and $\mathbf{V}$ are $P_{\mathbf{Y}}$ and $P_{\mathbf{Z}}$, respectively.

Assume that Alice sends $T$ independent messages $K_1, K_2, \ldots, K_T$ to Bob by a multiplex coding. Each $K_t$, $t = 1, 2, \cdots, T$, takes values in $\mathcal{K}_t \equiv \{1, 2, \cdots, M_t\}$, and satisfies that $\Pr\{K_t = k\} = 1/M_t$ for all $k \in \mathcal{K}_t$. The aim of the multiplex coding is to attain the following performance.

(A) Every $K_t$ must be transmitted to Bob within an arbitrarily small error probability.

(B) The perfect secrecy against Eve must be attained for each $K_t$, $t = 1, 2, \cdots, T$, individually.

Note that the above (B) does not require the perfect secrecy of the entire $(K_1, K_2, \ldots, K_T)$, which is usually required in the ordinary (i.e. non-multiplex) coding for wiretap channels. In the case of (B), some information about the combination of $(K_1, K_2, \ldots, K_T)$ may leak out. But, since $K_t$, $t = 1, 2, \cdots, T$, are assumed to be mutually independent, the combination has no meaning, and hence the individual perfect secrecy of $K_t$ is reasonable.

The tuple $(K_1, K_2, \ldots, K_T)$ is encoded by an encoder $\varphi_n$ to a codeword $X^n$, which is sent to Bob via the main channel $W^n$. In this paper, we consider the case that stochastic encoders can be used in addition to the case that only deterministic encoders can be used. For each $t$, Bob decodes $\widehat{K}_t$ by a decoder $\psi_n^t$ from the channel output $Y^n$. Let $\mathcal{D}_k^t$ be the decoding region of $k \in \mathcal{K}_t$. Then for each $t$, $\mathcal{D}_1^t, \mathcal{D}_2^t, \cdots, \mathcal{D}_{M_t}^t$ are mutually disjoint, and $\widehat{K}_t = k$ is decoded if $Y^n \in \mathcal{D}_k^t$. In the case that $\widehat{K}_t \neq K_t$ or $Y^n \notin \cup_{k=1}^{M_t} \mathcal{D}_k^t$, a decoding error occurs for the $t$-th message $K_t$.

We represent the above code by $\mathcal{C}_n(\{M_1, \cdots, M_T\}, \varphi_n, \psi_n)$, or $\mathcal{C}_n$ for short, and evaluate the performance in the following three viewpoints.

(a) Coding rate of each message $K_t$:

$$\frac{1}{n} \log M_t$$

(b) Average decoding error probability of each message $K_t$:

$$\varepsilon_n^t(\mathcal{C}_n) \equiv \frac{1}{M_t} \sum_{k=1}^{M_t} Q_k^t W^n(\overline{\mathcal{D}_k^t}), \tag{13}$$

where $\overline{\mathcal{D}_k^t}$ is the complement set of $\mathcal{D}_k^t$, and $Q_k^t(x^n)$ is the probability distribution of input $X^n$ generated by the encoder $\varphi_n$ for a message $K_t = k$. Note that $Q_k^t(x^n)$ depends on other messages $K_{t'}, t' \neq t$ (and some random numbers in the case of stochastic encoder). The probability distribution of the outputs $Y^n$ and $Z^n$ of the main and wiretap channels for the message $K_t = k$ are given by $Q_k^t W^n(y^n)$ and $Q_k^t V^n(z^n)$, respectively.

(c) Security measures :

$$I_n^t(\mathcal{C}_n) \equiv \frac{1}{n}I(K_t; Z^n) = \frac{1}{n}\frac{1}{M_t}\sum_{k=1}^{M_t} D(Q_k^t V^n \| P_{Z^n})$$

$$= \frac{1}{n}\frac{1}{M_t}\sum_{k=1}^{M_t} D(Q_k^t V^n \| \frac{1}{M_t}\sum_{k=1}^{M_t} Q_k^t V^n) \qquad (14)$$

$$d_n^t(\mathcal{C}_n) \equiv \frac{1}{M_t(M_t-1)}\sum_{k=1}^{M_t}\sum_{k'=1(k'\neq k)}^{M_t} \|Q_{k'}^t V^n - Q_k^s V^n\|_1$$

$$= \frac{1}{M_t(M_t-1)}\sum_{k=1}^{M_t}\sum_{k'=1(k'\neq k)}^{M_t}\sum_{z^n} |Q_{k'}^t V^n(z^n) - Q_k^t V^n(z^n)| \qquad (15)$$

Note that if $I_n^t(\mathcal{C}_n)$ is sufficiently small, then the message $K_t$ and the output $Z^n$ are almost independent, and hence Eve cannot obtain any information about $K_t$ from $Z^n$. On the other hand, $d_n^t(\mathcal{C}_n)$ is the security measure based on the variational distance. If $d_n^t(\mathcal{C}_n)$ is sufficiently small, then the output probability distribution $Q_k^t V^n$ is almost the same as $Q_{k'}^t V^n$ for any $k' \in \mathcal{K}_t$. This also means that Eve cannot obtain any information about $K_t$ from $Z^n$.

Now we define the achievable rates $R_t, t = 1, 2, \cdots, T$ for the multiplex coding as follows.

**Definition 3**

If there exists a sequence of code $\mathcal{C}_n$ that satisfies (16)–(19), then a rate-tuple $(R_1, R_2, \cdots, R_T)$ is called achievable for channels $(\mathbf{W}, \mathbf{V})$ in the sense of the security measure $I_n^t(\mathcal{C}_n)$. Furthermore, if there exists a sequence of code $\mathcal{C}_n$ that satisfies (16)–(18) and (20), then a rate-tuple $(R_1, R_2, \cdots, R_T)$ is called achievable for channels $(\mathbf{W}, \mathbf{V})$ in the sense of the security measure $d_n^t(\mathcal{C}_n)$.

$$\liminf_{n\to\infty} \frac{1}{n}\log\left(\prod_{t'=1}^{T} M_{t'}\right) \geq R_{\text{total}}, \qquad (16)$$

$$\limsup_{n\to\infty}\left[\frac{1}{n}\log\left(\prod_{t'=1}^{T} M_{t'}\right) - \frac{1}{n}\log M_t\right] \leq R_{\text{total}} - R_t, \ t = 1, 2, \cdots, T, \qquad (17)$$

$$\lim_{n\to\infty} \varepsilon_n^t(\mathcal{C}_n) = 0, \qquad t = 1, 2, \cdots, T, \qquad (18)$$

$$\lim_{n\to\infty} I_n^t(\mathcal{C}_n) = 0, \qquad t = 1, 2, \cdots, T, \qquad (19)$$

$$\lim_{n\to\infty} d_n^t(\mathcal{C}_n) = 0, \qquad t = 1, 2, \cdots, T, \qquad (20)$$

where the total rate $R_{\text{total}}$ is defined as

$$R_{\text{total}} = \sum_{t=1}^{T} R_t. \qquad (21)$$

**Remark 1** Note from (16) and (17) that if $(R_1, R_2, \cdots, R_T)$ is achievable, then it satisfies

$$\liminf_{n\to\infty}\frac{1}{n}\log M_n^t \geq R_t, \ t = 1, 2, \cdots, T \qquad (22)$$

because

$$
\begin{aligned}
R_t &\leq R_{\text{total}} - \limsup_{n\to\infty}\left[\frac{1}{n}\log\left(\prod_{t'=1}^{T} M_{t'}\right) - \frac{1}{n}\log M_t\right] \\
&\leq \liminf_{n\to\infty}\frac{1}{n}\log\left(\prod_{t'=1}^{T} M_{t'}\right) + \liminf_{n\to\infty}\left[-\frac{1}{n}\log\left(\prod_{t'=1}^{T} M_{t'}\right) + \frac{1}{n}\log M_t\right] \\
&\leq \liminf_{n\to\infty}\left[\frac{1}{n}\log\left(\prod_{t'=1}^{T} M_{t'}\right) - \frac{1}{n}\log\left(\prod_{t'=1}^{T} M_{t'}\right) + \frac{1}{n}\log M_t\right] \\
&= \liminf_{n\to\infty}\frac{1}{n}\log M_t.
\end{aligned}
$$

Therefore, if $(R_1, R_2, \cdots, R_T)$ is achievable, each message $K_t$ can be transmitted securely with at least the rate $R_t$. However, for any rate-tuple $(R_1, R_2, \cdots, R_T)$, there exists a sequence of code $\{\mathcal{C}_n\}$ that does not satisfy both equalities of (16) and (22). Such a case occurs if $\{\mathcal{C}_n\}$ satisfies

$$\liminf_{n\to\infty}\frac{1}{n}\log\left(\prod_{t'=1}^{T} M_{t'}\right) > \sum_{t'=1}^{T}\liminf_{n\to\infty}\frac{1}{n}\log M_{t'}. \tag{23}$$

In order to avoid this inconvenience, (17) is used instead of (22).

**Remark 2** Although (19) or (20) ensure the perfect secrecy of each message $K_t$, some information about the combination of $(K_1, K_2, \ldots, K_T)$ leaks out. Since $K_1, K_2, \ldots, K_T$ are mutual independent, the leaked information has no meaning. However, if an enemy gets a message $K_t$ by a method other than the output $\mathbf{Z}$ of the wiretap channel, the enemy may also get some information about other messages $K_{t'}, t' \neq t$. If Alice and Bob want to prevent the possibility of such attack, they must use the ordinary, i.e. non-multiplex, coding for wiretap channels.

**Definition 4**

Let $\mathcal{R}^I_{\text{det}}(\mathbf{W},\mathbf{V},T)$, $\mathcal{R}^d_{\text{det}}(\mathbf{W},\mathbf{V},T)$, $\mathcal{R}^I_{\text{sto}}(\mathbf{W},\mathbf{V},T)$, and $\mathcal{R}^d_{\text{sto}}(\mathbf{W},\mathbf{V},T)$ be the closures of achievable rate-tuples $(R_1, R_2, \cdots, R_T)$ for the main and wiretap channels $(\mathbf{W}, \mathbf{V})$. The subscript "det" represents the case that only deterministic encoders can be used while "sto" means that stochastic encoders including deterministic encoders can be used. Furthermore, the subscripts "$I$" and "$d$" stand for the cases that the security is measured by $I_n^t(\mathcal{C}_n)$ and $d_n^t(\mathcal{C}_n)$, respectively.

From the above definition, it holds obviously that for any $(\mathbf{W}, \mathbf{V})$ and any $T$,

$$\mathcal{R}^I_{\text{det}}(\mathbf{W},\mathbf{V},T) \subseteq \mathcal{R}^I_{\text{sto}}(\mathbf{W},\mathbf{V},T), \tag{24}$$

$$\mathcal{R}^d_{\text{det}}(\mathbf{W},\mathbf{V},T) \subseteq \mathcal{R}^d_{\text{sto}}(\mathbf{W},\mathbf{V},T). \tag{25}$$

Since the multiplex coding of plural messages is treated, we usually assume in this paper that $T \geq 2$. But, note that the case of $T = 1$ corresponds to the ordinary coding for wiretap channels.

In order to evaluate the above achievable rate regions, we first define two regions $\mathcal{R}_1(\mathbf{W},\mathbf{V},T)$ and $\mathcal{R}_2(\mathbf{W},\mathbf{V},T)$ as follows.

**Definition 5**

$$\mathcal{R}_1(\mathbf{W}, \mathbf{V}, T) = \{(R_1, R_2, \cdots, R_T) \,|\, \text{There exists an input probability distribution } P_{\mathbf{X}}$$
$$\text{that satisfies (27) and (28)}\}. \tag{26}$$

$$R_{\text{total}} \leq \underline{I}(P_{\mathbf{X}}, \mathbf{W}) \tag{27}$$
$$R_{\text{total}} - R_t \geq \overline{I}(P_{\mathbf{X}}, \mathbf{V}), \quad t = 1, 2, \cdots, T. \tag{28}$$

**Definition 6**

$$\mathcal{R}_2(\mathbf{W}, \mathbf{V}, T) = \{(R_1, R_2, \cdots, R_T) \,|\, \text{There exists an input probability distribution } P_{\widetilde{\mathbf{X}}} \text{ and}$$
$$\text{a test channel } \mathbf{U} \text{ with alphabets } \widetilde{\mathcal{X}} \to \mathcal{X} \text{ that satisfy}$$
$$(30) \text{ and } (31)\}. \tag{29}$$

$$R_{\text{total}} \leq \underline{I}(P_{\widetilde{\mathbf{X}}}, \mathbf{UW}) \tag{30}$$
$$R_{\text{total}} - R_t \geq \overline{I}(P_{\widetilde{\mathbf{X}}}, \mathbf{UV}), \; t = 1, 2, \cdots, T, \tag{31}$$

where $\widetilde{\mathcal{X}}$ is an arbitrary finite alphabet, and $\mathbf{UW}$ with $\widetilde{\mathcal{X}} \to \mathcal{Y}$ and $\mathbf{UV}$ with $\widetilde{\mathcal{X}} \to \mathcal{Z}$ are the cascade channels of ($\mathbf{U}$ and $\mathbf{W}$) and ($\mathbf{U}$ and $\mathbf{V}$), respectively.

For the above rate regions, the following theorems hold.

**Theorem 2** For any given $\mathbf{W}$, $\mathbf{V}$, and $T \geq 2$, $\mathcal{R}_{\text{det}}^I(\mathbf{W}, \mathbf{V}, T)$ and $\mathcal{R}_{\text{det}}^d(\mathbf{W}, \mathbf{V}, T)$ satisfy

$$\mathcal{R}_{\text{det}}^I(\mathbf{W}, \mathbf{V}, T) \supseteq \mathcal{R}_1(\mathbf{W}, \mathbf{V}, T), \tag{32}$$
$$\mathcal{R}_{\text{det}}^d(\mathbf{W}, \mathbf{V}, T) \supseteq \mathcal{R}_1(\mathbf{W}, \mathbf{V}, T), \tag{33}$$

respectively. Furthermore, if $\mathbf{V}$ satisfies (12), then it holds that

$$\mathcal{R}_{\text{det}}^d(\mathbf{W}, \mathbf{V}, T) = \mathcal{R}_1(\mathbf{W}, \mathbf{V}, T). \tag{34}$$

**Theorem 3** For any given $\mathbf{W}$, $\mathbf{V}$, and $T \geq 2$, $\mathcal{R}_{\text{sto}}^I(\mathbf{W}, \mathbf{V}, T)$ and $\mathcal{R}_{\text{sto}}^d(\mathbf{W}, \mathbf{V}, T)$ satisfy

$$\mathcal{R}_{\text{sto}}^I(\mathbf{W}, \mathbf{V}, T) \supseteq \mathcal{R}_2(\mathbf{W}, \mathbf{V}, T), \tag{35}$$
$$\mathcal{R}_{\text{sto}}^d(\mathbf{W}, \mathbf{V}, T) \supseteq \mathcal{R}_2(\mathbf{W}, \mathbf{V}, T), \tag{36}$$

respectively. Furthermore, if $\mathbf{V}$ satisfies (12), then it holds that

$$\mathcal{R}_{\text{sto}}^d(\mathbf{W}, \mathbf{V}, T) = \mathcal{R}_2(\mathbf{W}, \mathbf{V}, T). \tag{37}$$

**Remark 3** From (27), (28), (30) and (31), each $R_t$, $t = 1, 2, \cdots, T$ in Theorems 2 and 3 must satisfy that

$$R_t \leq \underline{I}(P_{\mathbf{X}}, \mathbf{W}) - \overline{I}(P_{\mathbf{X}}, \mathbf{V}), \tag{38}$$

$$R_t \leq \underline{I}(P_{\widetilde{\mathbf{X}}}, \mathbf{UW}) - \overline{I}(P_{\widetilde{\mathbf{X}}}, \mathbf{UV}), \tag{39}$$

respectively.

**Remark 4** We note from [11, Theorem 5 and the proof of Lemma 5] that in the case of $T = 1$, the secrecy capacity $C_S$ is given by

$$C_S = \sup_{P_{\widetilde{\mathbf{X}}}, \mathbf{U}: \overline{I}(P_{\widetilde{\mathbf{X}}}, \mathbf{UV}) = 0} \underline{I}(P_{\widetilde{\mathbf{X}}}, \mathbf{UW})$$

$$= \sup_{P_{\widetilde{\mathbf{X}}}, \mathbf{U}} \left[ \underline{I}(P_{\widetilde{\mathbf{X}}}, \mathbf{UW}) - \overline{I}(P_{\widetilde{\mathbf{X}}}, \mathbf{UV}) \right] \tag{40}$$

in both cases of the security measures $I_n^s(\mathcal{C}_n)$ and $d_n^t(\mathcal{C}_n)$. On the other hand, it holds from Definition 6 that for $T = 1$, $R_1 \leq \underline{I}(P_{\widetilde{\mathbf{X}}}, \mathbf{UW})$, $0 = \overline{I}(P_{\widetilde{\mathbf{X}}}, \mathbf{UV})$. Hence, it holds for $T = 1$ that

$$\mathcal{R}_{\text{sto}}^I(\mathbf{W}, \mathbf{V}, 1) = \mathcal{R}_{\text{sto}}^d(\mathbf{W}, \mathbf{V}, 1) = \mathcal{R}_2(\mathbf{W}, \mathbf{V}, 1) = [0, C_S]. \tag{41}$$

**Remark 5** Let $P_{\mathbf{X}}^*$ be the input probability distribution that can attain the channel capacity $C = \sup_{P_{\mathbf{X}}} \underline{I}(P_{\mathbf{X}}, \mathbf{W})$, i.e. $\underline{I}(P_{\mathbf{X}}^*, \mathbf{W}) = C$. Then, from Theorem 2 and using this $P_{\mathbf{X}}^*$ in (26), a rate-tuple $(R_1, R_2, \cdots, R_T)$ is achievable if it satisfies

$$R_{\text{total}} = C, \tag{42}$$

$$R_{\text{total}} - R_t \geq \overline{I}(P_{\mathbf{X}}^*, \mathbf{V}). \tag{43}$$

Note that in the case of $R_1 = R_2 = \cdots = R_T$, (43) holds for $T$ satisfying

$$T \geq \left\lceil \frac{\underline{I}(P_{\mathbf{X}}^*, \mathbf{W})}{\underline{I}(P_{\mathbf{X}}^*, \mathbf{W}) - \overline{I}(P_{\mathbf{X}}^*, \mathbf{V})} \right\rceil \geq \left\lceil \frac{C}{C_S} \right\rceil. \tag{44}$$

Therefore, the channel capacity can be attained as the total rate of the multiplex coding with an appropriate $T$, and each message $K_t$ can be individually protected perfectly against Eve.

## IV. Proofs

We first prove Theorem 2.

## A. Direct Part

The direct part, i.e. (32) and (33), can be proved in the same way as [11, the proof of Theorem 3], which uses the coding scheme introduced in [8].

In a code $\mathcal{C}_n$, the total number of codewords is given by $\prod_{t=1}^{T} M_t$. We generate every codeword independently with probability $P_{X^n}$. Then, let $x^n_{k_1,k_2,\cdots,k_T}$ be the codeword that corresponds to messages $K_t = k_t$, $t = 1, 2, \cdots, T$. The decoding regions $\mathcal{D}_{k_1,k_2,\cdots,k_T}$ for messages $K_t = k_t$, $t = 1, 2, \cdots, T$, are defined by

$$\mathcal{D}_{k_1,k_2,\cdots,k_T} \equiv \mathcal{A}(x^n_{k_1,k_2,\cdots,k_T}) \Big\backslash \bigcup_{(k'_1,k'_2,\cdots,k'_T) \neq (k_1,k_2,\cdots,k_T)} \mathcal{A}(x^n_{k'_1,k'_2,\cdots,k'_T}), \tag{45}$$

where $\mathcal{A}(x^n)$ is defined for a given real number $a$, which is determined later, as follows.

$$\mathcal{A}(x^n) \equiv \left\{ y^n \in \mathcal{Y}^n \mid \frac{W^n(y^n|x^n)}{P_{Y^n}(y^n)} > 2^{an} \right\} \tag{46}$$

We now evaluate the performance for the above random code ensemble. Note that for each message $K_t$ in the above code, all the other messages can be considered as a dummy message to keep the message $K_t$ secret from Eve. On the other hand, for the case of non-multiplex coding, Hayashi [11, Section IV] proved the coding theorem for wiretap channels by using the random code with message size $M$ and the dummy size $L$. For each $t$, letting

$$L_t \equiv \frac{\prod_{t'=1}^{T} M_{t'}}{M_t}, \tag{47}$$

the above random code ensemble coincides with Hayashi's random code ensemble with the message size $M = M_t$ and the dummy size $L = L_t$, and therefore the following Lemma holds from [11, the proof of Theorem 3].

**Lemma 1** The above random code ensemble satisfies for any real numbers $a$ and $b$ that for $t = 1, 2, \cdots, T$,

$$\mathrm{E}\varepsilon_n^t(\mathcal{C}_n) \leq \Pr\left\{ \frac{1}{n} \log \frac{W^n(Y^n|X^n)}{P_{Y^n}(Y^n)} < a \right\} + L_t \cdot M_t \cdot 2^{-an}$$

$$= \Pr\left\{ \frac{1}{n} \log \frac{W^n(Y^n|X^n)}{P_{Y^n}(Y^n)} < a \right\} + \left( \prod_{t=1}^{T} M_t \right) \cdot 2^{-an} \tag{48}$$

$$\mathrm{E}d_n^t(\mathcal{C}_n) \leq 2\left( 2\Pr\left\{ \frac{1}{n} \log \frac{V^n(Z^n|X^n)}{P_{Z^n}(Z^n)} > b \right\} + \sqrt{\frac{2^{bn}}{L_t}} \right) \tag{49}$$

$$\mathrm{E}I_n^t(\mathcal{C}_n) \leq \frac{1}{n}\eta(\delta_n) + \delta_n \cdot \log|\mathcal{Z}| + \frac{2^{bn}}{L_t}, \tag{50}$$

where E represents the expectation over the random code ensemble, $\eta(x) = -x \log x$ and $\delta_n$ is defined by

$$\delta_n = \Pr\left\{ \frac{1}{n} \log \frac{V^n(Z^n|X^n)}{P_{Z^n}(Z^n)} > b \right\}. \tag{51}$$

It holds from Markov's inequality [1] that for each $t$,

$$\Pr\left\{\left(\varepsilon_n^t(\mathcal{C}_n) \leq 3T \cdot \mathrm{E}\varepsilon_n^t(\mathcal{C}_n)\right)^c\right\} < \frac{1}{3T} \tag{52}$$

$$\Pr\left\{\left(d_n^t(\mathcal{C}_n) \leq 3T \cdot \mathrm{E}d_n^t(\mathcal{C}_n)\right)^c\right\} < \frac{1}{3T} \tag{53}$$

$$\Pr\left\{\left(I_n^t(\mathcal{C}_n) \leq 3T \cdot \mathrm{E}I_n^t(\mathcal{C}_n)\right)^c\right\} < \frac{1}{3T}, \tag{54}$$

where $(\mathcal{G})^c$ stands for the complement event of $\mathcal{G}$. Hence, among the random code ensemble, there exists a code satisfying all of (55)-(57).

$$\varepsilon_n^t(\mathcal{C}_n) \leq 3T \cdot \mathrm{E}\varepsilon_n^t(\mathcal{C}_n), \quad t = 1, 2, \cdots, T \tag{55}$$

$$d_n^t(\mathcal{C}_n) \leq 3T \cdot \mathrm{E}d_n^t(\mathcal{C}_n), \quad t = 1, 2, \cdots, T \tag{56}$$

$$I_n^s(\mathcal{C}_n) \leq 3T \cdot \mathrm{E}I_n^t(\mathcal{C}_n), \quad t = 1, 2, \cdots, T \tag{57}$$

Now we show by selecting parameters $M_t$, $a$, and $b$ adequately that for an arbitrarily given $P_\mathbf{X}$ and $\gamma > 0$, a rate-tuple $(R_1, R_2, \cdots, R_T)$ is achievable if it satisfies that

$$R_{\text{total}} \leq \underline{I}(P_\mathbf{X}, \mathbf{W}) - \gamma \tag{58}$$

$$R_{\text{total}} - R_t \geq \overline{I}(P_\mathbf{X}, \mathbf{V}) + \gamma, \quad t = 1, 2, \cdots, T. \tag{59}$$

Setting $M_t = 2^{nR_t}$, we have that

$$\prod_{t=1}^T M_t = 2^{nR_{\text{total}}}$$
$$\leq 2^{n(\underline{I}(P_\mathbf{X}, \mathbf{W}) - \gamma)}, \tag{60}$$

$$L_t = 2^{n(R_{\text{total}} - R_t)}$$
$$\geq 2^{n(\overline{I}(P_\mathbf{X}, \mathbf{V}) + \gamma)}. \tag{61}$$

By setting $a = \underline{I}(P_\mathbf{X}, \mathbf{W}) - \gamma/2$, and $b = \overline{I}(P_\mathbf{X}, \mathbf{V}) + \gamma/2$, we obtain that

$$\left(\prod_{t=1}^T M_t\right) \cdot 2^{-an} \leq 2^{-n\gamma/2} \tag{62}$$

$$\frac{2^{bn}}{L_t} \leq 2^{-n\gamma/2}. \tag{63}$$

Hence, from (48)-(50), (55)-(57), and Definition 1, it holds that

$$\lim_{n \to \infty} \varepsilon_n^t(\mathcal{C}_n) = 0, \tag{64}$$

$$\lim_{n \to \infty} d_n^t(\mathcal{C}_n) = 0, \tag{65}$$

$$\lim_{n \to \infty} I_n^t(\mathcal{C}_n) = 0, \tag{66}$$

which means that the rate-tuple $(R_1, R_2, \cdots, R_T)$ is achievable if it satisfies (58) and (59). Finally, since the above argument holds for any $\gamma > 0$, any rate-tuple in $\mathcal{R}_1$ is achievable in both senses of the security measures $d_n^t(\mathcal{C}_n)$ and $I_n^t(\mathcal{C}_n)$.

*B. Converse Part*

We will show that if the wiretap channel **V** satisfies (12) for any $P_\mathbf{X}$, then the following relation holds.

$$\mathcal{R}_{\text{det}}^d(\mathbf{W}, \mathbf{V}, T) \subseteq \mathcal{R}_1(\mathbf{W}, \mathbf{V}, T). \tag{67}$$

Let $(R_1, R_2, \cdots, R_T) \in \mathcal{R}_{\text{det}}^d(\mathbf{W}, \mathbf{V}, T)$. Then, there exists a sequence of code $\{\mathcal{C}_n\}$ that satisfies (16)–(18), and (20). Hence the code $\mathcal{C}_n$ satisfies that for any $\gamma > 0$ and any sufficiently large $n$,

$$\frac{1}{n} \log \left( \prod_{t'=1}^T M_{t'} \right) \geq R_{\text{total}} - \gamma \tag{68}$$

$$\frac{1}{n} \log \left( \prod_{t'=1}^T M_{t'} \right) - \frac{1}{n} \log M_t \leq R_{\text{total}} - R_t + \gamma, \quad t = 1, 2, \cdots, T. \tag{69}$$

Now let $X^n$ be the uniform random variable that takes values in the set of codewords $\{x_{k_1,k_2,\cdots,k_T}^n\}_{k_t \in \mathcal{K}_t, t=1,\cdots,T}$. Then, it holds from Verdú-Han's Lemma [5, Lemma 3.2.2] that for any $\gamma > 0$,

$$\sum_{t=1}^T \varepsilon_n^t(\mathcal{C}_n) \geq \Pr\left\{ \frac{1}{n} \log \frac{W(Y^n|X^n)}{P_{Y^n}(Y^n)} \leq \frac{1}{n} \log \left( \prod_{t=1}^T M_t \right) - \gamma \right\} - e^{-n\gamma}$$

$$\geq \Pr\left\{ \frac{1}{n} \log \frac{W(Y^n|X^n)}{P_{Y^n}(Y^n)} \leq R_{\text{total}} - 2\gamma \right\} - e^{-n\gamma} \tag{70}$$

Hence we conclude from (2) and (18) that $R_{\text{total}}$ must satisfy (27).

Next we prove (28) by considering the following variational distance for each $t$. Let $\widetilde{X}_k^n$ be the random variable that is the same as $X^n$ except that $K_t$ is fixed as $K_t = k$. Then, noting $P_{X^n} = \frac{1}{M_t} \sum_{k=1}^{M_t} P_{\widetilde{X}_k^n}$, we have that

$$\|P_{X^n} V^n - P_{\widetilde{X}_k^n} V^n\|_1 = \left\| \frac{1}{M_t} \sum_{k'=1}^{M_t} P_{\widetilde{X}_{k'}^n} V^n - P_{\widetilde{X}_k^n} V^n \right\|_1$$

$$= \frac{1}{M_t} \left\| \sum_{k'=1}^{M_t} (P_{\widetilde{X}_{k'}^n} V^n - P_{\widetilde{X}_k^n} V^n) \right\|_1$$

$$= \frac{1}{M_t} \left\| \sum_{k'=1(k' \neq k)}^{M_t} (P_{\widetilde{X}_{k'}^n} V^n - P_{\widetilde{X}_k^n} V^n) \right\|_1$$

$$\leq \frac{1}{M_t} \sum_{k'=1(k' \neq k)}^{M_t} \|P_{\widetilde{X}_{k'}^n} V^n - P_{\widetilde{X}_k^n} V^n\|_1$$

$$\leq \frac{1}{M_t - 1} \sum_{k'=1(k' \neq k)}^{M_t} \|P_{\widetilde{X}_{k'}^n} V^n - P_{\widetilde{X}_k^n} V^n\|_1. \tag{71}$$

Since the average of $\|P_{X^n} V^n - P_{\widetilde{X}_k^n} V^n\|_1$ for $k = 1$ to $M_t$ tends to zero asymptotically from (20) and (71), it must hold that for some $k$,

$$\lim_{n \to \infty} \|P_{X^n} V^n - P_{\widetilde{X}_k^n} V^n\|_1 = 0. \tag{72}$$

Noting that $\widetilde{X}_k^n$ is a random number which uniformly distributes over the set with $L_t = \left(\prod_{t'=1}^{T} M_{t'}\right)/M_t$ elements, $S_{\mathbf{X}}(\mathbf{V})$ must satisfy from Definition 2 and (69) that for any $\gamma > 0$,

$$S_{\mathbf{X}}(\mathbf{V}) \leq \limsup_{n \to \infty} \frac{1}{n} \log L_t \leq R_{\text{total}} - R_t + \gamma. \tag{73}$$

This means that if (12) holds for any $P_{\mathbf{X}}$, then (28) also holds.

In the sequel, any $(R_1, R_2, \cdots, R_T) \in \mathcal{R}_{\text{det}}^d(\mathbf{W}, \mathbf{V}, T)$ is included in $\mathcal{R}_1(\mathbf{W}, \mathbf{V}, T)$.

In the case of Theorem 3, we can use a test channel $\mathbf{U}$ with alphabets $\widetilde{\mathcal{X}} \to \mathcal{X}$ in the encoder $\varphi_n$ because stochastic encoders can be used. Hence, Theorem 3 can be proved in the same way as Theorem 2 by considering the cascade channels $(\mathbf{UW}, \mathbf{UV})$ instead of $(\mathbf{W}, \mathbf{V})$.

## V. STATIONARY MEMORYLESS WIRETAP CHANNELS

In this section, we consider the case that channels $\mathbf{W}$ and $\mathbf{V}$ are stationary memoryless channels with transition probability distribution $W$ and $V$, respectively. In this case, it holds that the spectral sup- and inf-mutual information rates are equal to the ordinary mutual information. Hence the following corollary holds from Theorems 2 and 3.

**Corollary 1**

If the channels $\mathbf{W}$ and $\mathbf{V}$ are stationary memoryless channels given by $W$ and $V$, respectively, it holds for $T \geq 2$ that

$$\mathcal{R}_{\text{det}}^I(\mathbf{W}, \mathbf{V}, T) \supseteq \mathcal{R}_1^*(W, V, T), \tag{74}$$

$$\mathcal{R}_{\text{sto}}^I(\mathbf{W}, \mathbf{V}, T) \supseteq \mathcal{R}_2^*(W, V, T), \tag{75}$$

where $\mathcal{R}_1^*(W, V, T)$ and $\mathcal{R}_2^*(W, V, T)$ are defined in Definition 7.

**Definition 7**

$$\mathcal{R}_1^*(W, V, T) = \{(R_1, R_2, \cdots, R_T) \,|\, \text{There exists an input probability distribution } P_{\widetilde{X}}$$
$$\text{that satisfies (78) and (79)}\}. \tag{76}$$

$$\mathcal{R}_2^*(W, V, T) = \{(R_1, R_2, \cdots, R_T) \,|\, \text{There exists an input probability distribution } P_X \text{ and}$$
$$\text{a test channel } U \text{ with alphabets } \widetilde{\mathcal{X}} \to \mathcal{X} \text{ that satisfy}$$
$$\text{(80) and (81)}\}. \tag{77}$$

$$R_{\text{total}} \leq I(P_X, W) \tag{78}$$

$$R_{\text{total}} - R_t \geq I(P_X, V), \quad t = 1, 2, \cdots, T. \tag{79}$$

$$R_{\text{total}} \leq I(P_{\widetilde{X}}, UW) \tag{80}$$

$$R_{\text{total}} - R_t \geq I(P_{\widetilde{X}}, UV), \quad t = 1, 2, \cdots, T, \tag{81}$$

where $\widetilde{X}$ is an auxiliary random variable over a finite alphabet $\widetilde{\mathcal{X}}$, [1] and $UW$ with $\widetilde{\mathcal{X}} \to \mathcal{Y}$ and $UV$ with $\widetilde{\mathcal{X}} \to \mathcal{Z}$ are the cascade channels of ($U$ and $W$) and ($U$ and $V$), respectively. The mutual informations are defined as $I(P_X, W) \equiv I(X;Y)$, $I(P_X, V) \equiv I(X;Z)$, $I(P_{\widetilde{X}}, UW) \equiv I(\widetilde{X};Y)$, and $I(P_{\widetilde{X}}, UV) \equiv I(\widetilde{X};Z)$ where the random variables make a Markov chain $\widetilde{X} \to X \to (Y, Z)$.

We note that $\delta_n$ defined in (51) goes to zero with an exponential order of $n$ in the case of stationary memoryless channels. Hence, even if we use

$$\widehat{I}_n^t(\mathcal{C}) \equiv I(K_t; Z^n) \tag{82}$$

instead of $I_n^t(\mathcal{C}_n)$ defined in (14) as a security measure, we can easily prove that the code shown in section IV-A also satisfies

$$\lim_{n \to \infty} \widehat{I}_n^t(\mathcal{C}_n) = 0. \tag{83}$$

Therefore, Corollary 1 holds for $\widehat{I}_n^t(\mathcal{C})$ similarly.

Furthermore, it is well known, e.g. refer [1], that the divergence $D(P_1 \| P_2)$ and the variational distance $\| P_1 - P_2 \|_1$ satisfies

$$D(P_1 \| P_2) \geq \frac{1}{2 \ln 2} \| P_1 - P_2 \|_1^2. \tag{84}$$

Hence, if a rate-tuple $(R_1, R_2, \cdots, R_T)$ is achievable for $\widehat{I}_n^t(\mathcal{C}_n)$, then the rate-tuple is also achievable for $d_n^t(\mathcal{C}_n)$. This means that $\mathcal{R}_{\det}^{\widehat{I}}(\mathbf{W}, \mathbf{V}, T) \subseteq \mathcal{R}_{\det}^d(\mathbf{W}, \mathbf{V}, T)$ and $\mathcal{R}_{\text{sto}}^{\widehat{I}}(\mathbf{W}, \mathbf{V}, T) \subseteq \mathcal{R}_{\text{sto}}^d(\mathbf{W}, \mathbf{V}, T)$.

By combining Theorems 2, 3 and the above facts, we obtain the following corollary.

**Corollary 2**

If the channels $\mathbf{W}$ and $\mathbf{V}$ are stationary memoryless channels given by $W$ and $V$, respectively, it holds that

$$\mathcal{R}_{\det}^d(\mathbf{W}, \mathbf{V}, T) \supseteq \mathcal{R}_{\det}^{\widehat{I}}(\mathbf{W}, \mathbf{V}, T) \supseteq \mathcal{R}_1^*(W, V, T) \quad \text{for } T \geq 2 \tag{85}$$

$$\mathcal{R}_{\text{sto}}^d(\mathbf{W}, \mathbf{V}, T) \supseteq \mathcal{R}_{\text{sto}}^{\widehat{I}}(\mathbf{W}, \mathbf{V}, T) \supseteq \mathcal{R}_2^*(W, V, T) \quad \text{for } T \geq 2. \tag{86}$$

Furthermore, if the wiretap channel $V$ is full-rank, then the following equalities also hold.

$$\mathcal{R}_{\det}^d(\mathbf{W}, \mathbf{V}, T) = \mathcal{R}_{\det}^{\widehat{I}}(\mathbf{W}, \mathbf{V}, T) = \mathcal{R}_1^*(W, V, T) \quad \text{for } T \geq 2 \tag{87}$$

$$\mathcal{R}_{\text{sto}}^d(\mathbf{W}, \mathbf{V}, T) = \mathcal{R}_{\text{sto}}^{\widehat{I}}(\mathbf{W}, \mathbf{V}, T) = \mathcal{R}_2^*(W, V, T) \quad \text{for } T \geq 2 \tag{88}$$

**Remark 6** In the case of $T = 1$, we note from Remark 4 that for stochastic encoders, all the achievable rate regions for three security measures coincide with $\mathcal{R}_2^*(W, V, 1)$, and the maximum $R_1 \in \mathcal{R}_2^*(W, V, 1)$ is equal to the secrecy capacity $C_S$ determined by Csiszár and Körner [3]. Hence, it holds that

$$\mathcal{R}_{\text{sto}}^d(\mathbf{W}, \mathbf{V}, 1) = \mathcal{R}_{\text{sto}}^I(\mathbf{W}, \mathbf{V}, 1) = \mathcal{R}_{\text{sto}}^{\widehat{I}}(\mathbf{W}, \mathbf{V}, 1) = \mathcal{R}_2^*(W, V, 1) = [0, C_S], \tag{89}$$

---

[1]$|\widetilde{\mathcal{X}}|$, the cardinality of $\widetilde{\mathcal{X}}$, can be bounded by $|\widetilde{\mathcal{X}}| \leq |\mathcal{X}| + 1$. Refer [3] for more details.

where $C_S$ is given by
$$C_S = \max_{\widetilde{X} \to X \to (Y,Z)} [I(\widetilde{X}; Y) - I(\widetilde{X}; Z)]. \tag{90}$$